\pdfoutput=1
\documentclass{PoS}

\title{$B$ and $D$ Meson Decay Constants}

\ShortTitle{$B$ and $D$ Meson Decay Constants}

\author{C. Bernard,$^1$ C. DeTar,$^2$ M. Di Pierro,$^3$ A. X. El-Khadra,$^4$ R. T. Evans,$^4$ 
E. D. Freeland,$^5$ E. Gamiz,$^4$ S.~A.~Gottlieb,$^6$ U. M. Heller,$^7$ J. E. Hetrick,$^8$ 
A. S. Kronfeld,$^9$ J. Laiho,$^1$, $^9$ L. Levkova,$^2$ 
\speaker{P. B. Mackenzie},$^9$  
J. Simone,$^9$ R. Sugar,$^{10}$ D. Toussaint,$^{11}$ and R. S. Van~de~Water$^9$ 
\\
$^1$Washington University, 
$^2$University of Utah, 
$^3$DePaul University, 
$^4$University of Illinois, 
$^5$The School of the Art Institute of Chicago, 
$^6$Indiana University, 
$^7$American Physical Society, 
$^8$University of the Pacific, 
$^9$Fermi National Accelerator Laboratory, 
$^{10}$University of California, Santa Barbara, 
$^{11}$University of Arizona \\ 
        E-mail: \email{mackenzie@fnal.gov}
}

\author{ Fermilab Lattice and MILC Collaborations}

\abstract{We present an update of our calculations of the decay
constants of the $D$, $D_s$, $B$, and $B_s$ mesons in unquenched
$2+1$ flavor  QCD.
We use the MILC library of improved staggered gauge ensembles at 
lattice spacings 0.09, 0.12, and 0.15 fm,
clover heavy quarks with the Fermilab normalizations,
and improved staggered light valence quarks.
}

\FullConference{The XXVI International Symposium on Lattice Field Theory\\
		 July 14-19 2008\\
		 Williamsburg, Virginia, USA}

\begin{document}

\section{Introduction}

In 2005, combined work by the Fermilab Lattice and MILC Collaborations
\cite{Aubin:2005ar} determined the value of the $D_s$ decay constant
$f_{D_s}$ to around 10\% before it had been determined to that accuracy
by experiment.  When the subsequent experimental determination agreed to
within one sigma,
we claimed that as a successful prediction. As lattice calculations
become increasingly accurate, of course,
at some point we do not expect perfect agreement
 between the Standard Model and experiment.
With sufficient precision, the effects of Beyond-the-Standard-Model physics
will start to show up in low energy measurements.
We do not know what that precision will be, so
we must be cautious in interpreting deviations between theory and experiment.

Since then, we have increased the precision of our calculations.
Our result for $f_{D_s}$ remains about 10\% below the experimental result, and
with the increased precision of theory and experiment, no longer agrees 
to within one sigma with experiment, as we describe in this paper.
Further, earlier this year new results on the $\pi$, $K$, $D$, and $D_s$
decay constants appeared from the HPQCD Collaboration \cite{Follana:2007uv}.
They used a new lattice fermion method, Highly Improved Staggered Quarks, or 
``HISQ'' fermions,
which allowed them to calculate all four decay constants with nearly identical 
methods.
They found very good agreement with experiment for the $\pi$ and $K$
decay constants.
Their value for the $D$ decay constant was subsequently confirmed by 
CLEO \cite{:2008sq}.
For $f_{D_s}$, they also found a result around 10\% below experiment, but with 
improved precision.
Instead of agreement between theory and experiment, there is now a greater
than three sigma discrepancy.
This is the only quantity in  lattice QCD phenomenology
with staggered fermions in which such a clear 
disagreement has arisen between theory and experiment, 
so a puzzle has developed.  Precise calculations
with other lattice methods are of great interest.

\section{Calculations}

We are finishing a reanalysis of the existing data for our calculations
of $f_D$, $f_{D_s}$, $f_B$, and $f_{B_s}$
that is reducing some of our largest uncertainties.
We are also preparing for new runs this year with four times the statistics.
Our calculations are done with improved staggered (``asqtad'') light
quarks \cite{Lepage:1998vj,Orginos:1998ue}, 
and clover/Fermilab \cite{ElKhadra:1996mp} ${\cal O}(a)$ improved heavy quarks.
We use the MILC $2+1$ flavor library of unquenched gauge configurations
\cite{Aubin:2004fs},
with lattice spacings of around 0.15, 0.12, and 0.09 fm (the so-called
coarser, coarse, and fine ensembles).
The masses of light sea-quarks range between $0.6 m_s$ and $0.1 m_s$.
On each of the eleven ensembles, we use from eight to twelve
 partially quenched valence quark masses,
ranging from around $m_s$ to $0.1 m_s$.

The decay constants are defined by
\begin{equation}
\left<0|A_\mu|H_q(p)\right>=if_{H_q}p_\mu.
\end{equation}
The combination decay amplitude
\begin{equation}
\phi_{H_q}=f_{H_q}\sqrt{M_{H_q}}
\end{equation}
can be obtained from the correlators
\begin{eqnarray}
C_0(t)     &=& \left<O^\dagger_{H_q}(t) O_{H_q}(0)\right>,  \\
C_{A_4}(t)  &=& \left<A_4(t)  O_{H_q}(0) \right>.
\end{eqnarray}
The current normalizations are obtained from
\begin{equation}
Z^{Qq}_{A_4} = \rho^{Qq}_{A_4}\sqrt{Z^{QQ}_{V_4}Z^{qq}_{V_4}},
\end{equation}
where $Z^{QQ}_{V_4}$ and $Z^{qq}_{V_4}$
 are determined nonperturbatively and the remaining
(perturbatively calculated) short distance corrections
in the deviation of $\rho^{Qq}_{A_4}$ from 1
 are no more than 0.6\%.

\begin{figure}[htb]
\center{\includegraphics[width=0.63\textwidth]{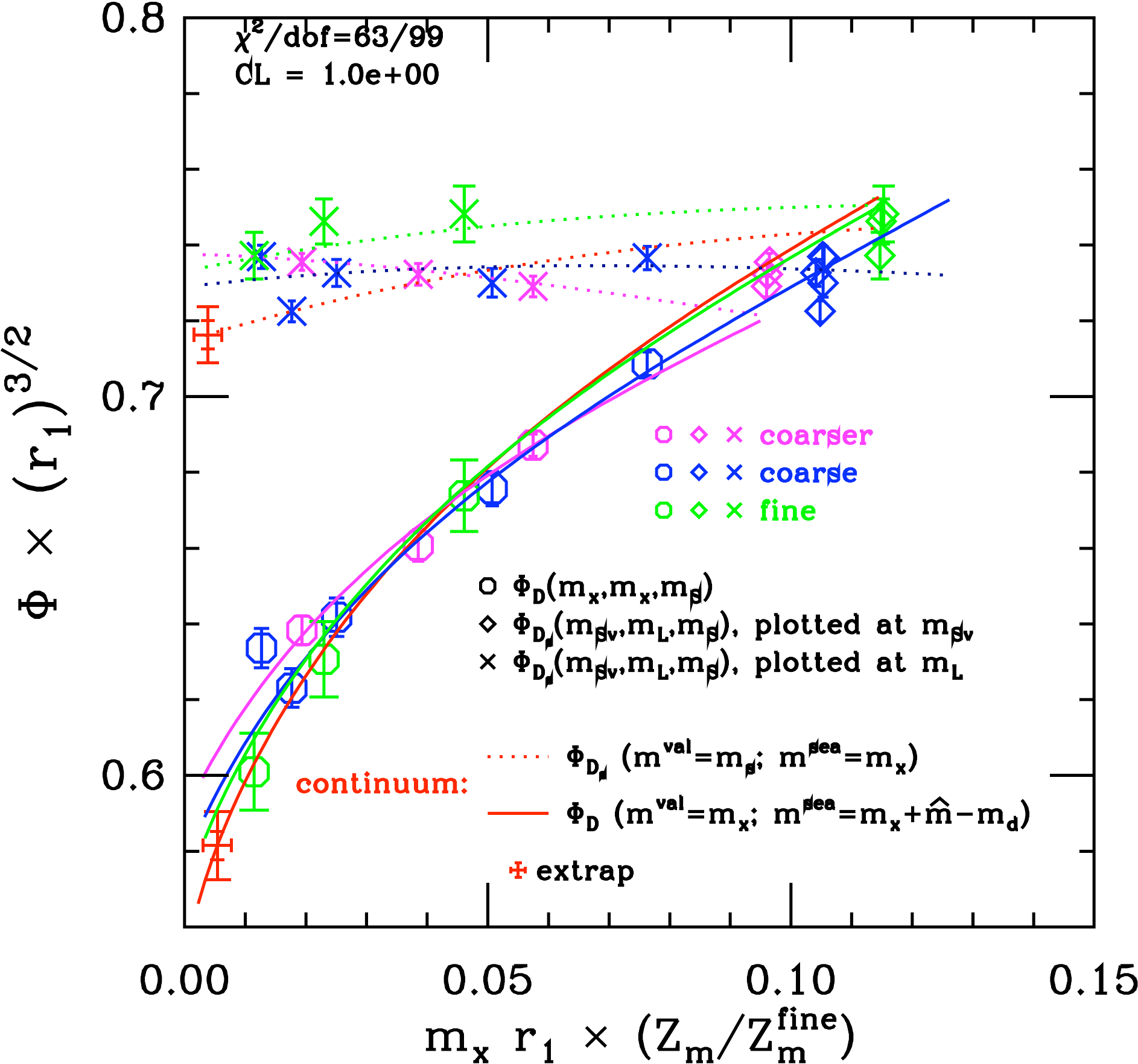}}
\caption{The leptonic decay amplitudes $f\sqrt{M}$ for the $D$ and
$D_s$ mesons, extrapolated to the chiral limit.
Units are in terms of the heavy-quark potential parameter, $r_1$.
}
\vspace{-5pt}
\label{figs/phiDextrap.pdf}
\end{figure}

Figure~\ref{figs/phiDextrap.pdf} shows the extrapolation of the $D$ and $D_s$ 
leptonic decay amplitudes to the physics light quark limit.
(Units are in terms of the heavy-quark potential parameter $r_1$.)
For $\phi_D$ (octagons), we show only those (fully unquenched) points for
 which the light valence and sea masses are equal
to $m_x$, the mass on the abscissa.  For $\phi_{D_s}$, we keep both 
the strange sea mass ($m_S$) and the strange valence mass
($m_{Sv}$) fixed to their simulated values, and plot either as a function 
of the up/down sea mass $m_L$ (crosses), or at $m_{Sv}$ (diamonds).
The chiral extrapolations make use of all the partially quenched data 
in addition to the points shown.
 
\begin{figure}[htb]
\center{\includegraphics[width=0.63\textwidth]{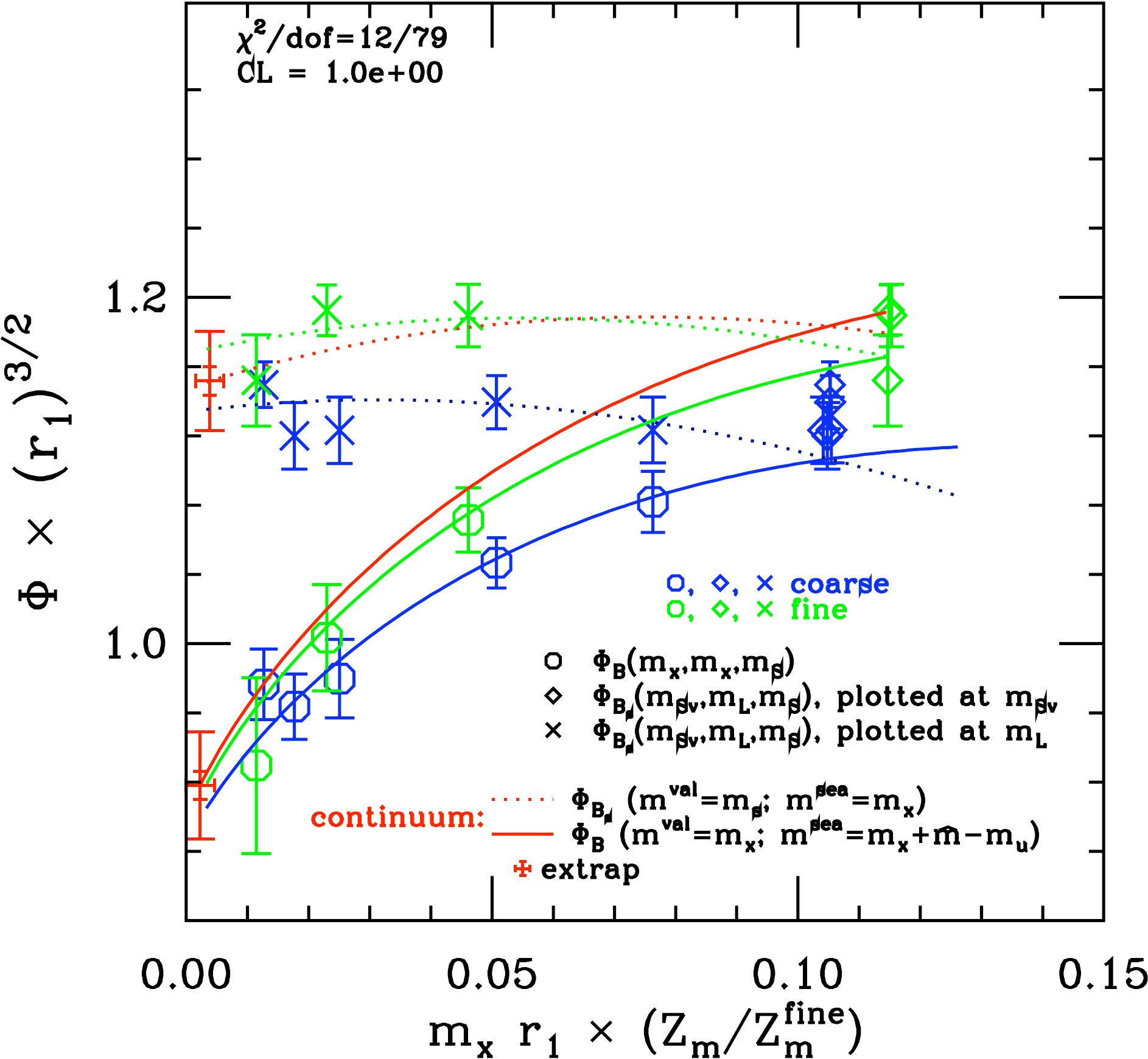}}
\caption{The same as for Fig.~1, but for the
$B$ and $B_s$ mesons.
}
\vspace{-5pt}
\label{figs/phiBextrap.pdf}
\end{figure}

The $m_q$ dependence is much stronger for the $D$ than the $D_s$, as expected,
since in the $D_s$ it affects only the sea quarks and not the valence quarks.
The slope is larger in the continuum limit, because
taste breaking effects tend to suppress the dependence on the quark mass at
finite $a$.
Figure \ref{figs/phiBextrap.pdf} shows the same thing for the $B$
and $B_s$ decay amplitudes, with a qualitatively similar picture.

\begin{figure}[htb]
\center{\includegraphics[width=0.63\textwidth]{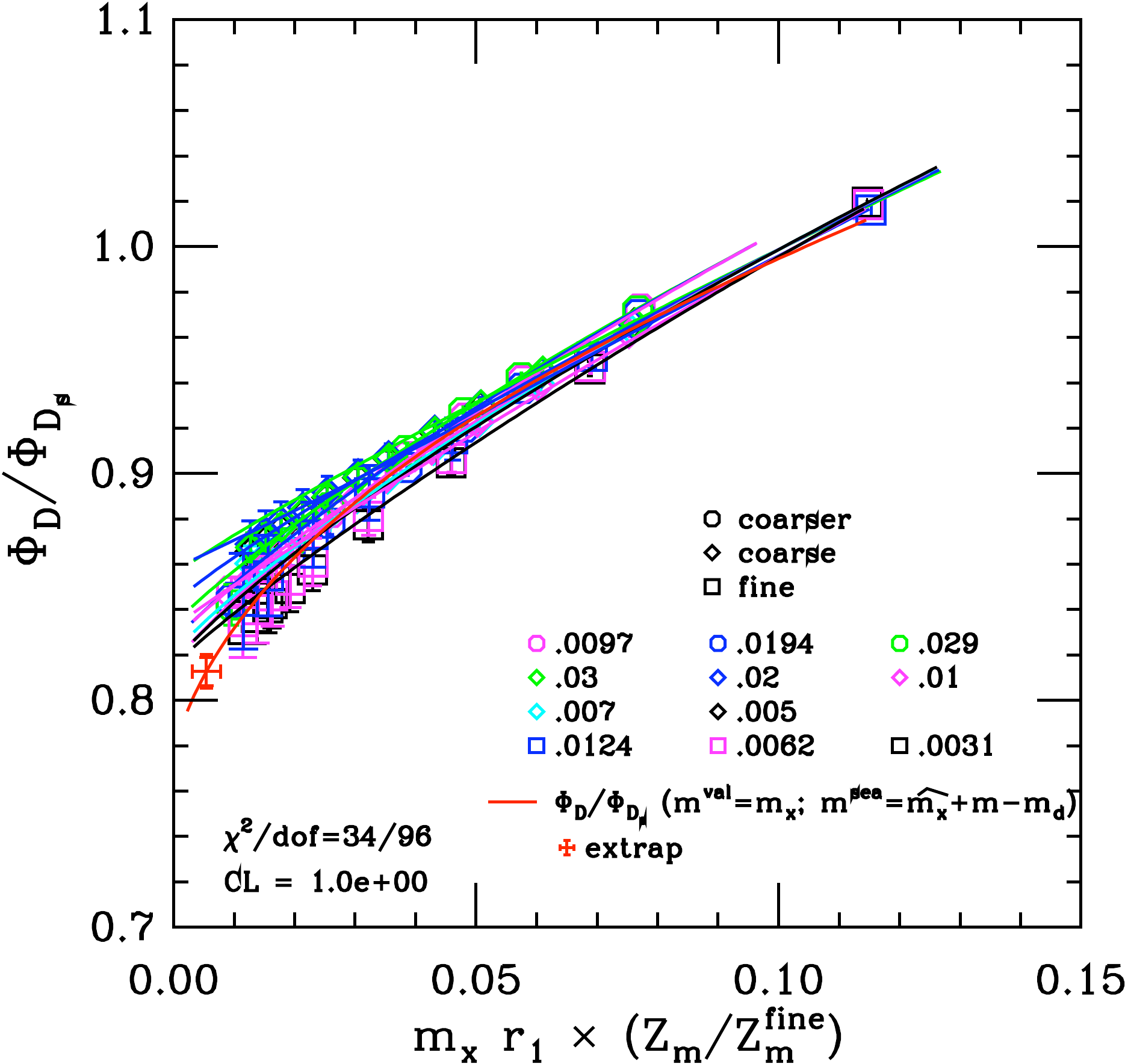}}
\caption{The ratio of the $D$ and $D_s$ meson leptonic decay amplitudes, extrapolated
to the chiral limit.
}
\vspace{-5pt}
\label{figs/ratiofit.pdf}
\end{figure}

Figure~\ref{figs/ratiofit.pdf} shows the extrapolation of the ratio $\phi_D/\phi_{D_s}$
to the chiral limit.
The slope is strongest in the continuum limit (red line and cross), as 
expected.

\section{Results}

Table~1
shows the uncertainty budgets for the
$D_s$, $D$, $B_s$, and $B$ meson decay amplitudes $\phi_M$, and for the
ratios $R_D\equiv \phi_D/\phi_{D_s}$ and $R_B\equiv \phi_B/\phi_{B_s}$.
The three largest uncertainties in our previous results were statistics,
heavy quark discretization, and the heavy quark mass.
The statistical error in the $D$ and $D_s$ decay amplitudes has been reduced
this year through an improvement in analysis method, and without additional data.
We are currently incorporating into the chiral and continuum extrapolation
fits a term for the heavy quark discretization which we expect to substantially
reduce the uncertainty from this source.
The last uncertainty that is  large is due to the input heavy quark mass, and
 will be removed with a more careful determination of this quantity.

\begin{table}[b]
\center{
\caption{Uncertainty budgets in per cent for the leptonic decay amplitudes
 $\phi_{D_s}$, $\phi_D$, $\phi_{B_s}$, and $\phi_B$, and for the ratios 
$R_D\equiv \phi_D/\phi_{D_s}$ and $R_B\equiv \phi_B/\phi_{B_s}$.
}
\vspace{.2 mm}
\begin{tabular}{|l|r|r|r|r|r|r|}
\hline
  &$\phi_{D_s}$  &$\phi_{D_d}$  &$R_D$  &$\phi_{B_s}$  &$\phi_{B_d}$  & $R_B$ \\
\hline
Statistics  &1.0  &1.5  &1.0  &2.5  &3.4  &2.2  \\  
\hline
Inputs $r_1, m_s, m_l$  &1.4  &2.1  &0.6  &1.8  &2.5  &0.6  \\  
Inputs $m_b$ or $m_c$   &2.7  &2.7  &0.1  &1.1  &1.1  &0.1  \\  
$Z$  &1.4  &1.4  &<0.1  &1.4  &1.4  &<0.1  \\  
Higher order $\rho_{A_4}$  &0.1  &0.1  &<0.1  &0.4  &0.4  &<0.1  \\  
Heavy q disc.  &2.7  &2.7  &0.3  &1.9  &1.9  &0.2  \\  
Light q disc. \& $\chi$ extr.  &1.2  &2.6  &1.6  &2..0  &2.4  &2.4  \\  
$V$  &0.2  &0.6  &0.6  &0.2  &0.6  &0.6  \\  
\hline
Total systematic  &4.5  &5.3  &1.8  &3.8  &4.4  &2.6  \\  
\hline
\end{tabular}
}
\label{uncertainties}
\end{table}

We obtain for the decay constants

\begin{eqnarray}
  f_D       &=&  207(11)\ {\rm MeV},  \\
  f_{D_s}    &=&  249(11)\ {\rm MeV},  \\
  f_B       &=&  195(11)\ {\rm MeV},  \\
  f_{B_s}    &=&  243(11)\ {\rm MeV},  
\end{eqnarray}
and for the ratios
\begin{eqnarray}
f_D/f_{D_s}    &=&  0.833(19),    \\
f_B/f_{B_s}    &=&  0.803(28).    
\end{eqnarray}

\begin{figure}[b]
\includegraphics[width=\textwidth]{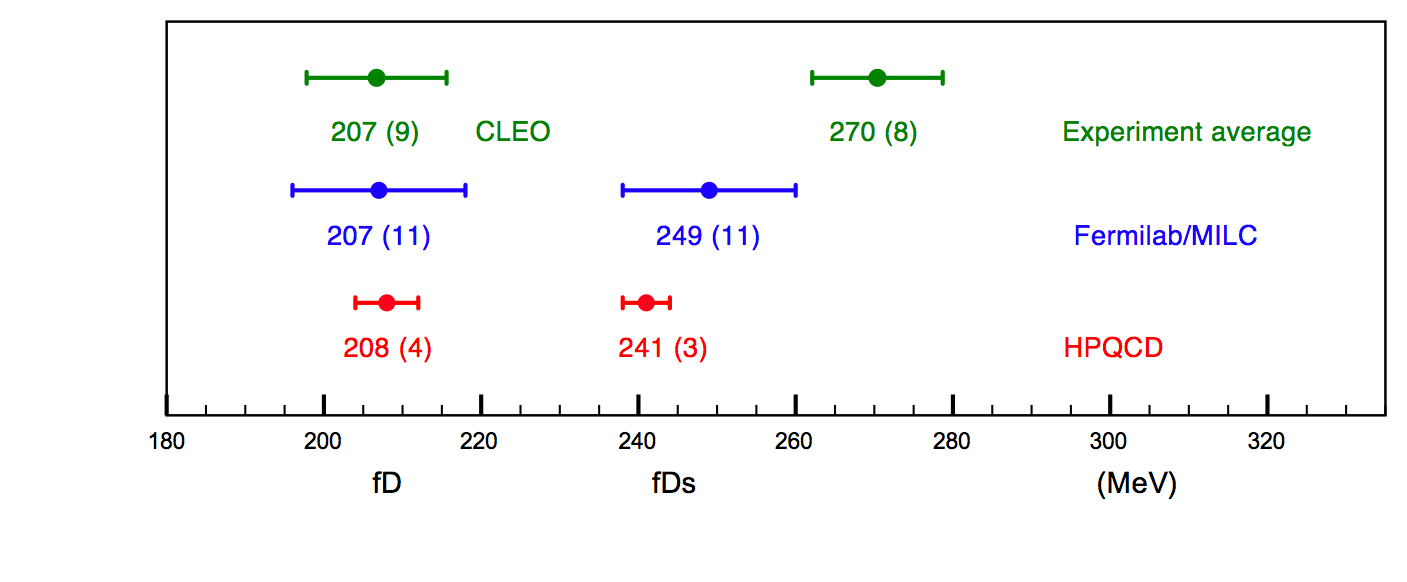}
\caption{Comparison of $f_D$ and $f_{D_s}$ with the calculations
of HPQCD and with experiment.
}
\vspace{-5pt}
\label{figs/fD(s).png}
\end{figure}

In Figure~\ref{figs/fD(s).png}, we compare our results for
$f_D$ and $f_{D_s}$ with the calculations
of HPQCD \cite{Follana:2007uv} 
and with experiment \cite{:2008sq,Rosner:2008yu}.
For $f_D$, there is very good agreement between experiment, HPQCD, and the 
Fermilab/MILC result.
For $f_{D_s}$, there is
\begin{itemize}
\item agreement between HPQCD and Fermilab/MILC,
\item 1.6 $\sigma$ disagreement between Fermilab/MILC and experiment, and
\item 3.5 $\sigma$ disagreement between HPQCD and experiment.
\end{itemize}

\begin{figure}[htb]
\includegraphics[width=\textwidth]{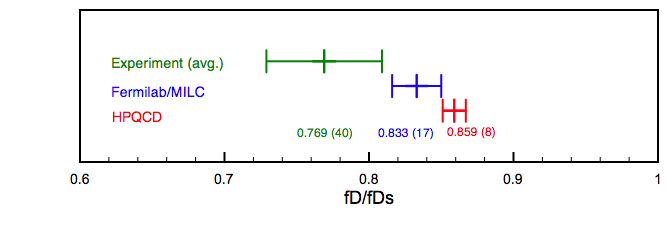}
\caption{Results for $f_D/f_{D_s}$ compared with the calculations of HPQCD
and with experiment.
}
\vspace{-5pt}
\label{figs/fDs-fD.png}
\end{figure}

Many uncertainties cancel in the ratio $f_D/f_{D_s}$,
so we examined this quantity to see if it could enhance the significance of the 
discrepancy between our results and experiment.
For now, looking at $f_D/f_{D_s}$ doesn't sharpen the picture.
In Figure~\ref{figs/fDs-fD.png}, we show our results for $f_D/f_{D_s}$ 
compared with the calculations of HPQCD
and with experiment.
There is a slight disagreement between HPQCD and FNAL/MILC in the ratio, even
though $f_D$ and $f_{D_s}$ agree within one sigma.
Further, the experimental uncertainties are independent.
They add in quadrature, increasing the size of the experimental uncertainty and
decreasing the significance of any discrepancy.

New results for $f_D$ and $f_{D_s}$ recently appeared 
from ETMC using twisted-mass fermions \cite{Blossier:2008dj}.
They obtained $f_D=205\pm 10$ MeV and $f_{D_s}=248\pm 9$ MeV, 
which is in accord with the staggered determinations.
They present a thorough uncertainty analysis, although we would quibble with
their use of two rather than three light sea quarks without the inclusion
of an uncertainty estimate for that approximation.
Based on the difference between our unquenched and quenched calculations 
of $f_{D_s}$ (249 MeV vs. 213 MeV) \cite{ElKhadra:1997hq}, 
we might have guessed a possible uncertainty
of 5\% from leaving out one of the three light sea quarks.  (We see charm sea
quarks as a different story, since $m_c\sim 1/a$ at our lattice spacings,
 and the dynamical effects
of $c$ quarks  are for the most part above the cut-off.)

Three sigma discrepancies between experiment and the Standard Model have 
occasionally appeared and then disappeared before, but
the discrepancy in $f_{D_s}$ is hard to 
understand.  The uncertainty is dominated by experimental statistical error,
and three sigma statistical fluctuations are very rare.
One can double the theory error, and still have a three sigma
discrepancy.
 To explain the discrepancy as a theory error, one would have to find
a mistake in the theory analysis of $f_{D_s}$
 whose correction would not affect the correct prediction of $f_D$.
It is hard to imagine such a mistake.
The calculations of $f_D$ and $f_{D_s}$ are almost identical.
The only difference is that $f_{D_s}$ should be somewhat easier, in that it doesn't
require an extrapolation to the physical light quark masses.
It may be premature to draw ultimate conclusions about the discrepancy, but the
 result is puzzling enough that Kronfeld and Dobrescu have investigated possible
new-physics explanations for the discrepancy~\cite{Dobrescu:2008er}.

\section{Outlook}

We are in the process of reanalyzing our existing data, in which we hope to bring
down several of our largest uncertainties.
New runs are starting with quadruple the current statistics
and at smaller lattice spacings,
which we expect to help with several of the uncertainties.
Comparison of $f_{D_s}$ in theory and experiment remains a puzzle.
This is the only known instance in which lattice QCD with staggered fermions
seems to clearly fail to reproduce the Standard Model.
This provides a good target of opportunity for calculations with other
lattice fermion methods.

\end{document}